\documentstyle[aps,twocolumn]{revtex}
\newcommand{\be}{\begin{equation}}
\newcommand{\ee}{\end{equation}}
\newcommand{\bea}{\begin{eqnarray}}
\newcommand{\eea}{\end{eqnarray}}
\newcommand{\bit}{\begin{itemize}}
\newcommand{\eit}{\end{itemize}}
\newcommand{\ben}{\begin{enumerate}}
\newcommand{\een}{\end{enumerate}}

\begin{document}
\title
{Knight Shift Anomalies in Heavy Electron Materials}

\author{E. Kim$^1$, M. Makivic$^{1,2}$ and D. L. Cox$^1$\\
$^1$Department of Physics, Ohio State University, Columbus, OH 43210\\
$^2$Northeast Parallel Architectures Center, Syracuse, NY 13244}
\date{\today}
\maketitle
\setlength{\baselineskip}{12pt}

\widetext

\begin{abstract}
We calculate non-linear Knight Shift $K$ vs. susceptibility $\chi$ anomalies
for Ce ions possessing
local moments in metals.  The ions are
modeled with the Anderson Hamiltonian and studied within
the non-crossing approximation (NCA).
The $K-vs.- \chi$ non-linearity diminishes with decreasing Kondo temperature
$T_0$ and nuclear spin-
local moment separation.
Treating the Ce ions as an incoherent array in CeSn$_3$,
we find excellent agreement with the observed Sn $K(T)$ data.
\smallskip
\end{abstract}
\bigskip

\narrowtext

\vspace{.4in}

The origin of Knight shift anomalies in metals with localized
moments that undergo the Kondo effect
has been a subject of great interest in the condensed matter
community over a
period of nearly twenty five
years\cite{heeger,bs74,gz74,i76,i78,mv75,mb79,m85,cjk87,ph88,lm94}.
The central physical concept is that the
many body screening
cloud surrounding a Kondo impurity site should give rise to an
anomalous temperature
dependent Knight shift at nuclear sites due to the coupling of the
local moment to the
nuclear spin through the screening cloud\cite{heeger,gz74,i76,i78,cjk87,ph88}.
This would be manifest in a
deviation from a
linear relation of the Knight shift $K$ to the magnetic susceptibility
$\chi$ below the Kondo temperature $T_0$.
Another way to say this is that in the absence of an
anomaly, the contribution $K(\vec r,T)$ from a
local moment at
distance $\vec r$ from the nucleus can be written as $f(\vec r)\chi(T)$.
This factorization
does not hold
if there is an anomaly (instead $K(\vec r,T) = f(\vec r,T)\chi(T)$ due to
the temperature dependent polarization cloud).
The classic experiments by Boyce and Slichter\cite{bs74}
on the low Kondo temperature ($T_0\simeq 10K$) alloy Fe:Cu displayed no
evidence for this
anomalous
Knight shift behavior.  In contrast,
pronounced Knight shift
anomalies have been observed in the concentrated heavy
electron materials CeSn$_3$\cite{mv75,m85} and YbCuAl\cite{mb79,m85},
which have been described as
Kondo lattice
systems with $T_0\simeq 400K$.  In view
of the Boyce-Slichter
result, the question is raised whether these anomalies represent a
coherent effect of the periodic
lattice rather than a single ion effect. However, recent experiments on
the
proposed quadrupolar Kondo alloy\cite{c87,sm91}
Y$_{1-x}$U$_x$Pd$_3$ demonstrate that
for concentrations
of 0.1-0.2 there are pronounced non-linearities in the Y Knight shift
for sufficiently large
distances away from the U ion\cite{lm94}.

In this paper, we present some results from our systematic theoretical
studies of the Knight
shift in heavy electron materials.
 We find, as expected from
earlier analytic theories, that the magnitude of the Knight shift
anomaly is reduced with
decreasing Kondo temperature
and decreasing distance between local
moment and impu-
 \newpage\hfill\\*[3.5cm]
rity
sites. Ours is the first study to display this effect in a realistic
model
calculation.
We find that the Knight
shift
anomaly of CeSn$_3$ can be accounted for by incoherent,
single ion physics.  In particular, by
carrying out a full lattice sum on CeSn$_3$, we obtain an excellent one
parameter fit of
the calculated Knight shift to the experimental one, with a small low
temperature deviation
possibly due to the over-simplifications of our model.
We conjecture that coherence effects may explain some of the small numerical
disagreement.

The application of NMR in Kondo systems received impetus from
Heeger\cite{heeger}
who suggested that the anomalous cloud was detectable in Knight shift
measurements.
Essentially, the oscillatory conduction
spin density $\vec s(\vec r)$ induces a local moment-nuclear moment
interaction.
In the limit $T_0\to 0$, this interaction ${\cal I}(\vec r)$ is
well described by the second order perturbation theory
Ruderman-Kittel-Kasuya-Yosida
expression
($\vec s(r)\sim {\cal I}(\vec r) \sim cos(2k_Fr)/r^3$).
Ishii\cite{i76,i78} confirmed
that for an $S_I=1/2$ pure-spin Kondo impurity (with no charge fluctuations)
an anomalous conduction spin density cloud sets in beyond the
Kondo screening length $\xi_K=\hbar v_F/k_BT_0$, where $v_F$ is the Fermi
velocity.
Inside this radius, conventional
temperature independent RKKY oscillations dominate of the kind observed
by Boyce and Slichter.
Outside the screening length, at $T=0$,
the anomalous term will dominate also with an RKKY form
but an amplitude of order $D$, the conduction bandwidth, compared with
$D(N(0)J)^2$ for the Ruderman-Kittel term, where $N(0)$ is the conduction
electron density of states at the Fermi energy and $J$ the conduction electron-
local moment exchange coupling.  Ishii did not calculate the explicit
temperature
dependence of this structure, but did anticipate that it would vanish above the
Kondo scale.
Scaling analysis confirmed the asymptotic factorization of the
Knight shift for short distance and low temperature\cite{cjk87}.
A possible understanding of the
Boyce and Slichter results, then, is the Cu nuclei they sampled were
at distances $r<<\xi_K$ from the Fe ions.

In an Anderson model treatment of the problem, charge fluctuations are allowed.
Consider, for example, a Ce ion with nominally one $f$ electron giving rise to
the local moment.
The ensuing Kondo effect is best understood in terms of a mixing of the empty
configuration
with one in which the Ce moment is screened by a superposition of conduction
hole states.
This opens up new possibilities for detecting the anomalous screening cloud.
One
interpretation of the Boyce-Slichter results is that due to subtleties of spin
conservation in the singlet ground state the spin cloud is unobservable, while
the
anomalous charge cloud is observable through electronic field gradient
measurements
at atomic nuclei\cite{gz74}.

We shall focus on the
second possibility, that the anomalous spin cloud now has a finite amplitude
within $r<\xi_K$ which is proportional to the weight of $f^0$
in the ground state.  This anomaly scales approximately linearly with $T_0$, so
the cloud
will be virtually undetectable in small $T_0$ systems, but may be observable in
high $T_0$ systems.  Our results support this picture as we shall explain in
more
detail.

Our starting point is an Anderson model Hamiltonian.  We shall first discuss
the
situation for Ce$^{3+}$ and Yb$^{3+}$ ions, and write down the model only for
the Ce case.  For a single Ce site at the origin, the model is
\begin{equation}
{\cal H} = {\cal H}_c + {\cal H}_f + {\cal H}_{cf} +{\cal H}_z
\end{equation}
with
\begin{equation}
{\cal H}_c = \sum_{\vec k\sigma} \epsilon_k c^{\dagger}_{\vec k\sigma}c_{\vec
k\sigma}
\end{equation}
the conduction band term for electrons with a broad featureless density of
states of
width $D$, taken to be Lorentzian here for convenience;
with
\begin{equation}
{\cal H}_f = \sum_{jm} \epsilon_{fj} |f^1jm><f^1jm|
\end{equation}
where $j=5/2,7/2$ indexes the angular momentum multiplets of the Ce ion having
azimuthal quantum numbers $m$, with
$\epsilon_{f5/2}=-2 eV$,$\epsilon_{f7/2}=\epsilon_{f5/2}+\Delta_{so}=-1.7eV$
(we take the $f^0$ configuration at
zero energy);
with
\begin{equation}
{\cal H}_{cf} = {\sum_{\vec kjm\sigma}} [V_{\vec kj\sigma m} c^{\dagger}_{\vec
k\sigma}|f^0><f^1jm| +
h.c.]
\end{equation}
where $V_{\vec kj\sigma m} =VY_{3m-\sigma}(\hat
k)<3m-\sigma,1/2\sigma|jm>/\sqrt{N_s}$,
$V$ being the
one particle hybridization strength and $N_s$ the number of sites;
with
\begin{equation}
{\cal H}_z =- \mu_BH_z[2\sum_{\vec k\sigma} \sigma n_{\vec k,\sigma} -
\sum_{jm} g_jm|f^1jm><f^1jm|]
\end{equation}
being the Zeeman energy of the electronic system for a magnetic field $H_z$
applied along
the z-axis.  In addition to this, we must add a term coupling the nuclear spin
system
to the conduction electrons, which we take to be of a simple contact form $\sim
vec I(\vec r)\cdot
\vec S(\vec r_I)$ for each nuclear spin $\vec I(\vec r)$ at position $\vec r$
with $\vec S(\vec
R)$ the conduction spin density at the nuclear site, and a nuclear Zeeman term.
In terms of the parameters, the Kondo scale characterizing the low energy
physics is
given by
\begin{equation}
k_BT_0 = D({D\over \Delta_{so}})^{4/3}
({\Gamma\over\pi|\epsilon_{f5/2}|})^{1/6}
\exp({\pi\epsilon_{f5/2}\over 6\Gamma})
\end{equation}
where the hybridization width $\Gamma = \pi N(0)V^2$.

We treat the Anderson hamiltonian with the non-crossing approximation (NCA), a
self-
consistent diagrammatic perturbation theory discussed at length in the paper of
Bickers
{\it et al.}\cite{bcw}. This approximation is controlled by the large orbital
degeneracy
of the Ce ground state. It does show pathological behavior (due to the
truncation of the
diagrammatic expansion) for a temperature scale $T_p<<T_0$ in this conventional
Anderson
model, provided the $f^1$ occupancy $n_f \ge 0.7$.  In practice, this is not a
problem
for $N\ge 4$ as shown in Ref. \cite{bcw}, in that comparison of NCA results
with exact
thermodynamics from the Bethe-Ansatz shows agreement at the few percent level
above $T_p$.
Hence, this is a reliable method for our purposes.

The approach in the NCA is to write a propagator for each
ionic state of the Ce site (i.e., $f^1 j=5/2,7/2$ and $f^0$ in the present
model),
solve self-consistent integral equations to second order in the hybridization
for the ionic
propagator self-energies, and then to calculate physical properties as
convolutions of these propagators.
To evaluate the Knight shift we employ the lowest order diagram coupling
nuclear spin
to Ce magnetic moment, as shown
in Fig. 1.
The full expression corresponding to this diagram is cumbersome and shall be
presented in detail elsewhere.
To the extent that the dynamics of the empty orbital
can be neglected, this expression factorizes into a nearly temperature
independent RKKY interaction (modified due to
the spin-orbit coupling and anisotropic hybridization from the original form)
times
the $f$-electron susceptibility.  Thus, no anomaly results from the diagram in
this
limit. In this limit, the susceptibility in the diagram corresponds to the
the leading order estimate used in Ref. \cite{bcw} to compare with
exact Bethe-Ansatz results.

However, for $T=0$, the empty orbital
propagator may be written in an approximate two-pole form, one with amplitude
$1-Z$,
$Z=\pi k_BT_0/6\Gamma$, centered near zero energy, and one with amplitude $Z$
centered
at $\sim \epsilon_{f5/2}-k_BT_0$ which reflects the anomalous ground state
mixing due
to the Kondo effect.  The first term gives conventional RKKY oscillations
modulo the
anisotropy and altered range dependence induced by the $m,\hat k$ dependence of
the hybridization.
The amplitude of the second term goes to zero above the Kondo
temperature.  It is this term which may be traced to the anticipated anomalous
Knight
shift, and within such a two-pole approximation may be seen to be finite within
$\xi_K$,
have a stronger distance dependence in that regime, but possess an amplitude of
order
$Z$ only within this distance regime.  Beyond $\xi_K$, the amplitude is of
order 1/$N$
and
the shape of the spin oscillations
is the same as that found from the high frequency pole of the empty orbital
propagator.

The diagram of Fig. 1 has been studied with the NCA previously\cite{ph88}, but
only
for the spin 1/2 model with infinite Coulomb repulsion, and for a limited
parameter regime (very low $T_0$ values) and short distances ($r<<\xi_K)$. In
consequence,
no strong evidence was found for a Knight shift anomaly in this previous work.

Our numerical procedure, briefly, consists of solving the NCA integral for the
Anderson Hamiltonian specified above on a logarithmic mesh with order 600
points chosen
to be centered about the singular structures near the ground state energy
$E_0\sim
\epsilon_{f5/2}$.  We then feed the self-consistent propagators for the empty
and doubly
occupied orbitals into the convolution integrals obtained from the diagram of
Fig. 1,
which allows for evaluation of the Knight shift at arbitrary angle and distance
from
the nuclear site.  It is convenient to take the nuclear
site as the origin in this case leading to phase factors $e^{-i\vec k\cdot R}$
in the hybridization Hamiltonian ${\cal H}_{cf}$, where $\vec r$ is the
nuclear-Ce site separation.
These factors give the oscillations and position space angular dependence in
$K$.
All contributions from the Ce susceptibility are included.

To demonstrate the dependence of the $K-vs.-\chi$ anomaly as a function of
Kondo
scale $T_0$,
we have evaluated the diagram of Fig. 1 for a single
local moment placed at a fixed distance $r=3.3k_F^{-1}=3.3\AA$  for our choice
of
$k_F=1\AA^{-1}$
and angle of 0$^o$ with respect
the quantization axis.  We have tuned the Kondo scale  by holding fixed all
parameters
except the hybridization.  The Knight shift is scaled to a susceptibility by
matching
at high temperatures, and the susceptibility units are scaled by a fit of one
calculation to the data for CeSn$_3$
assuming $D=3 eV$.  The results are shown in Fig. 2(a).  Clearly, as the Kondo
scale
is reduced, the magnitude of the deviation between a linear $K vs. \chi$
relation is
systematically reduced.  Also, as the separation $r$ is decreased (c.f., Fig.
2(b))
the magnitude of the $K-vs.-\chi$ non-linearity diminishes.
Very similar results are obtained for Yb compounds (the
Yb$^{3+}$ ion has a lone 4f hole and our procedure decribes these with a simple
particle hole transformation, which we shall discuss in detail in a subsequent
publication).

To assess the relevance of this single site physics to the periodic compound
CeSn$_3$
we have performed a lattice sum about a given Sn nucleus from all surrounding
Ce ion.  We assume the Knight shift contribution
of each ion to be described by this single site physics,
known to be
a good approximation at high temperatures where the ions are incoherent with
one another,
and known to provide a very accurate description of the thermodynamics in many
cases.
We carried out the sum to several hundred shells about the Sn site, obtaining
good convergence at all calculated temperatures.
We fixed the parameters by fitting the susceptibility data of CeSn$_3$.
The result is shown in Fig. 3, where we have scaled the Knight shift by an
intermediate
range temperature
match to the susceptibility (note that the NMR data of Ref. \cite{mv75} extends
only to room temperature). There are several notable features here: (i) the
calculated
amplitude of our Knight shift prior to scaling is actually negative, which
implies that
the fit is sensible only if the assumed contact coupling between conduction and
nuclear
spins is negative, which actually makes sense because the Sn nucleus should
dominantly
couple through core polarization; (ii) the magnitude of the anomaly goes in the
right
direction and begins at the right temperature to agree with the experimental
anomaly\cite{mv75},
though the quantitative value is slightly too high--given the highly
oversimplified
conduction band we are employing, we don't regard this as a serious defect in
the
calculation; (iii) the magnitude of the anomalous contribution from the Ce
sites does
go down with distance from the Sn nucleus--the theoretical data which most
closely match
those of experiment are taken at the fixed distance $r=3k_F^{-1}=3\AA$ and
angle of
0$^o$. However, the anomaly is then still surprisingly large even at this short
distance.
While we cannot rule out that the discrepancy reflects our oversimplified band
structure,
we conjecture that lattice coherence effects may drive our calculated lattice
sum in
this direction, because the repeated scattering of conduction electrons off of
the Ce
ions (now in every unit cell) has the effect of reducing the screening
length\cite{millee}.
We intend to study this idea further within a local approximation ($d=\infty$
expansion)
to the lattice model. We note that the high temperature deviation of Knight
shift and
susceptibility may be realistic and is related to the different couplings to
ground
and excited spin orbit multiplets.

In summary, we have computed Knight shift anomalies within a realistic model
for Ce ions
in metals for the first time.  We find that the magnitude of the non-linearity
scales
down in size as the Kondo scale is diminished or the nuclear moment-local
moment separation
is reduced.  Modeling the Ce ions incoherently in CeSn$_3$, we find that
summing over all
nattice sites the resulting Sn Knight shift agrees quite well with experiment.
The remaining
quantitative discrepancies may be due to a combination of our oversimplified
band structure
for the conduction states together with lattice coherence effects.

This research was supported by a grant
from the U.S. Department of Energy, Office of Basic Energy Sciences,
Division of Materials
Research.  We acknowledge many useful conversations over the years
with D.E. MacLaughlin
and H. Lukefahr.

{ \bf Figure 1}. Feynman diagram of coupling between Ce local moment
and nuclear spin
in the infinite $U$ Anderson model.
The dashed line represents the singly occupied ($f^1$) state and wavy line
the empty orbital ($f^0$) state.  The solid line represents the conduction
electron.
The dot-dashed   line is a propagator for the nuclear
spin states and $\vec{H}$ is the extenal magnetic field.

{\bf Figure 2}  (a) Calculated Knight shift $K(T)$ vs. susceptibility $\chi(T)$
for a single Ce site
at $k_Fr=3.33$ from a nuclear moment and angle
$\Theta =0$.   Fixing the $f$-level energy $\epsilon_{5/2}=-2$eV, and the
spin orbit splitting $\Delta_{so}=0.3eV$, the hybridization
is varied to illustrate the dependence of the
nonlinearity on the magnitude of $T_0$.
The diagram of Fig. 1 is used to calculate $K(T)$.
The magnitude of the non-linearity diminishes as $T_0$ is reduced.
The theoretical Knight shift has been shifted by a common offset and
scale factor  to match the susceptibility.
(b) Calculated Knight shift $K(T)$ for a single Ce site vs. susceptiblity
$\chi(T)$
for varying separation with the Kondo scale used to fit the CeSn$_3$ $\chi(T)$
(see Fig. 3).  For each plot the angle is held at $\theta$ with respect
to the nuclear moment-Ce axis.  The magnitude of the non-linearity diminishes
as $k_Fr$ is
reduced.  The theoretical Knight shifts have been shifted by offset
and scale factors to match the susceptibilities; this does not affect
the relative size of the anomaly.

{\bf Figure 3} Temperature dependence of Sn Knight Shift $K(T)$ and Ce
$\chi(T)$ (both
calculated and experimental results\cite{mv75}) for CeSn$_3$. The theoretical
$K(T)$ is
calculated using the diagram of Fig. 1 and with $T_0$ chosen to fit the
experimental
$\chi(T)$ data. A full (incoherent) lattice sum is carried
out over several hundred shells of atoms.

\end{document}